\documentclass[prb,twocolumn,superscriptaddress, showpacs,preprintnumbers,amsmath,amssymb,superscriptaddress]{revtex4}

\pdfoutput=1
\usepackage{color}
\usepackage{graphicx}
\usepackage{dcolumn}
\usepackage{bm}
\usepackage{subfig}

\newcommand{\ket}[1]{| #1 \rangle}
\newcommand{\bra}[1]{\langle #1 |}

\newcommand{\proj}[1]{| #1 \rangle \langle #1 |}

\newcommand{\op}[1]{ \hat{\sigma}_{#1} }

\def\beq{\begin{equation}}
\def\eeq{\end{equation}}
\def\be{\begin{equation}}
\def\ee{\end{equation}}
\def\ben{\begin{eqnarray}}
\def\een{\end{eqnarray}}
\def\ba{\begin{aligned}}
\def\ea{\end{aligned}}

\def\bibfnamefont{ }
\def\bibnamefont{ }

\def\ul{ \bf }
\renewcommand{\ul}{}
\renewcommand{\ul}[1]{#1}

\begin{document}

\title{Entanglement witnessing and quantum cryptography with non-ideal ferromagnetic detectors}

\author{Waldemar~K{\l}obus}
\affiliation{Faculty of Physics, Adam Mickiewicz University, 61-614 Pozna\'{n}, Poland}

\author{Andrzej~Grudka}
\affiliation{Faculty of Physics, Adam Mickiewicz University, 61-614 Pozna\'{n}, Poland}

\author{Andreas~Baumgartner}
\affiliation{Department of Physics, University of Basel, Klingelbergstrasse 82, CH-4056 Basel, Switzerland}

\author{Damian~Tomaszewski}
\affiliation{Institute of Molecular Physics, Polish Academy of Science, 60-179 Pozna\'{n}, Poland}

\author{Christian~Sch\"{o}nenberger}
\affiliation{Department of Physics, University of Basel, Klingelbergstrasse 82, CH-4056 Basel, Switzerland}

\author{Jan~Martinek}
\email{martinek@ifmpan.poznan.pl}
\affiliation{Institute of Molecular Physics, Polish Academy of Science, 60-179 Pozna\'{n}, Poland}

\date{\today}

\begin{abstract}

We investigate theoretically the use of non-ideal ferromagnetic contacts as a mean to detect quantum entanglement of electron spins in transport experiments. We use a designated entanglement witness and find a minimal spin polarization of $\eta > 1/\sqrt{3} \approx 58 \%$ required to demonstrate spin entanglement. This is significantly less stringent than the ubiquitous tests of Bell's inequality with $\eta > 1/\sqrt[4]{2}\approx 84\%$. In addition, we discuss the impact of decoherence and noise on entanglement detection and apply the presented framework to a simple quantum cryptography protocol. Our results are directly applicable to a large variety of experiments.
\end{abstract}


\maketitle

\section{Introduction}

Quantum entanglement is a fundamental aspect of Nature and an essential resource in quantum information protocols.
Entanglement between freely propagating photons has been extensively investigated over decades. One of many important steps is the electrically driven generation of entangled photons on demand demonstrated recently.\cite{Salter_Nature465_2010} In such experiments entanglement is detected in correlation experiments with standard large-efficiency polarization beam splitters as an important element. In solid state devices the electron spin degree of freedom can in principle be used to encode qubits.\cite{Loss_DiVincenzo_PRA57_1998} A straightforward way to detect electron spins is to measure the electrical current through ferromagnetic materials. Since most materials show only a partial spin polarization, we face the question of what the consequences of non-ideal detectors are for entanglement detection and different quantum information schemes. In addition, in a solid state environment the generation of two-particle entanglement and its detection are rather difficult due to the strong coupling to the environment.

The controlled generation of electron spin entanglement came into reach with theoretical proposals\cite{Recher_Loss_PRB63_2001, lesovik_martin_blatter_EPJB01} and experimental realizations where a superconductor is used as a
source of correlated electrons in a process known as Cooper pair splitting.\cite{Hofstetter_Nature_2009,
Herrmann_PRL104_2010, Hofstetter_Baumgartner_PRL107_2011, Schindele_Baumgartner_PRL109_2012, Das_Heiblum_NatureComm_2012,Braunecker13} Similarly, also a quantum dot in a singlet state might act as a source of spin-entangled electrons.\cite{Saraga_Loss_PRL90_2003,Legel07} Proposals to detect entanglement between electron spin states in a solid state environment are usually based on a test of Bell's inequality.
Such inequalities not only test quantum entanglement, but more generally the quantum description of Nature, and thus are more restrictive than required to make a statement about entanglement alone. In other words: Bell's inequalities are not optimal entanglement witnesses.

Here we report the use of various entanglement witnesses based on ferromagnetic contacts and compare it to a test of Bell's inequality in the presence of noise and decoherence.\cite{Larsson_PRA57_1998} The analysis can be extended easily to other quantum information tasks in a variety of systems and we briefly discuss the example of a simple quantum cryptography application.

In a ferromagnetic material the density of states for the two spin species $\uparrow$ (up) and $\downarrow$ (down) are different. For the electron transport in the linear regime the properties at the Fermi energy are relevant and one defines the spin polarization
\ben
\eta = \frac{\nu_{\uparrow}(E_{F})-\nu_{\downarrow}(E_{F})}{\nu_{\uparrow}(E_{F})+\nu_{\downarrow}(E_{F})}.
\een
Spin information is translated into a charge signal by the fact that the probability of an $\uparrow$ electron to enter the magnetic material is different than for $\downarrow$ electron. However, most magnetic materials show non-ideal spin polarization with $ \eta < 1$, which leads to ambiguous information about the spin states, i.e. the spin is wrongly detected with the probability $1-\eta$. The effect of this measurement error directly leads to the question of what the lower limits of spin detection efficiency are for a given task, e.g. to detect quantum entanglement or to perform a quantum cryptography protocol.

The paper is organized as follows:
In Sec.~\ref{werner} we review Werner states, a mixture of maximally entangled state and white noise.
Section~\ref{det} describes the non-ideal ferromagnetic detection process, while in Sec.~\ref{dep} we show the equivalence of ferromagnetic non-ideal detection with ideal transmission channels and ideal detection with a noisy depolarizing channel. In Sec.~\ref{s:EW} we construct an entanglement witnesses and calculate the required detector efficiencies.
In Sec.~\ref{bell} we compare these results to a test of Bell's inequalities and in Sec.~\ref{crypto} we apply the formalism to a simple quantum cryptography task.
In Sec.~\ref{noisy} and Sec.~\ref{deph} we discuss the impact of white noise as well as spin relaxation and dephasing, respectively.
We summarize our results in Sec.~\ref{dis} and propose a road map for the detection of electron spin entanglement.

\section{Partially entangled Werner states} \label{werner}

\begin{figure}[b]
\begin{center}
\subfloat[\,]{\label{figeta}
\includegraphics[width=0.35\textwidth]{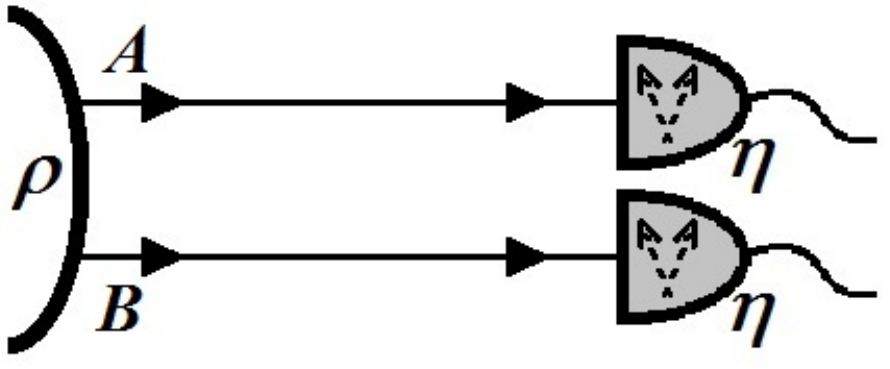}}
\qquad
\subfloat[\,]{\label{figeps}\includegraphics[width=0.35\textwidth]{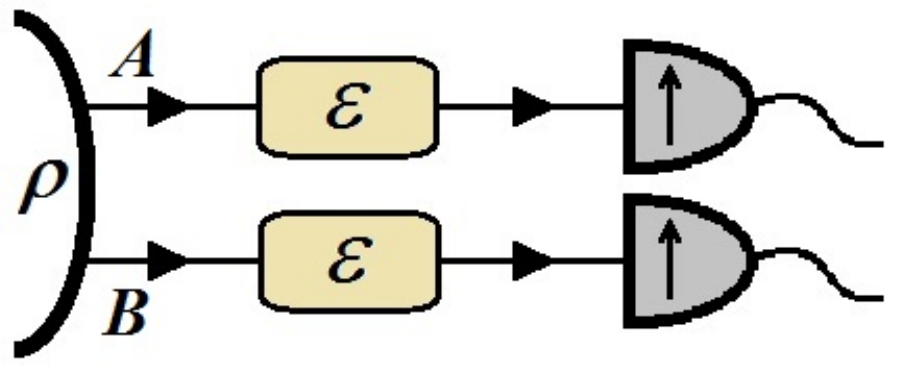}}
\end{center}
\caption{Schematic of a source of entangled electrons (left) connected to two conducting channels for two subsystems (electrons) $A$ and $B$, each with a ferromagnetic detector. The subfigures depict two equivalent ways of considering a non-ideal measurement: (a) a quantum state $\rho$ is measured using non-ideal ($\eta<1$) ferromagnetic detectors, or (b) each particle passes through a noisy depolarizing channel ($\mathcal{E}$) which transforms $\rho$, which is then measured with ideal detectors ($\eta = 1$).}
\label{setup}
\end{figure}

Figure~\ref{setup} shows a generic entanglemet detection scheme, where a source emits pair-wise entangled electrons into two (e.g. spatially) separated transport channels, each connected to a ferromagnetic detector. For example, the two particles can be in a singlet (Bell) state
\ben\label{singlet}
\ket{\Psi^-}_{AB} = \frac{1}{\sqrt2}(\ket{\uparrow}_A\otimes\ket{\downarrow}_B - \ket{\downarrow}_A\otimes\ket{\uparrow}_B),
\een
where $A$ and $B$ denote the two subsystems.

In real experiments the emitted quantum state is affected by external sources of noise and decoherence discussed later, with the effect that the original entangled state evolves into a mixed state. The form of this new state depends on the nature of the interaction with the environment, but it is most commonly modeled as 'white noise' in the form of a Werner state,\cite{Wer} i.e. a mixture of a maximally entangled state and white noise:
\ben\label{mixed}
\rho^{W}_{AB}(\lambda) = \lambda \proj{\Psi^-}_{AB} + (1-\lambda)\frac{I}{4},
\een
where $0\leq\lambda\leq1$ is the visibility parameter and $I$ the identity operator. 

\begin{figure}[t]
  \includegraphics[width=0.45\textwidth]{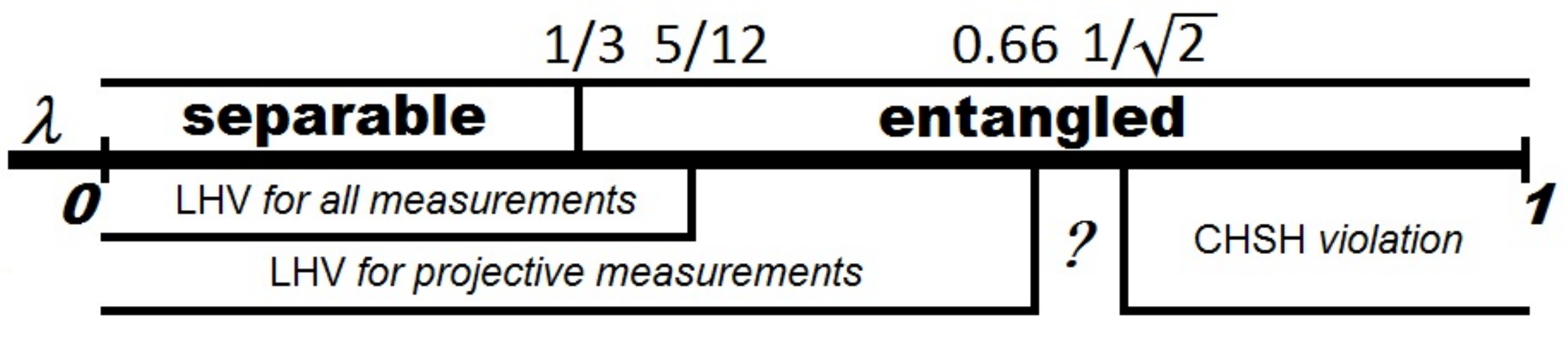}
  \caption{Regimes of entanglement, LHV models and nonlocality for Werner states with visibility parameter $\lambda$.}
  \label{figwer}
\end{figure}

Werner states have been studied intensely in the field of quantum information theory and to investigate the relation between quantum entanglement and nonlocality.\cite{Wer}
Different quantum information tasks require different minimal visibilities, which is depicted schematically in Fig~\ref{figwer}. Conclusive methods to decide whether a given two-qubit state is entangled was provided in Refs.~[\onlinecite{AP, Hor.sep}]. Applied to a Werner state, one finds that it is entangled for $\lambda > 1/3$. It was shown that there exist entangled Werner states for which a local hidden variable (LHV) model cannot be excluded,\cite{Wer} and that an LHV model exists for all measurements for $\lambda \leq 5/12$,\cite{Bar} and for all projective measurements for $\lambda \leq 0.6595$.\cite{AGT}

Werner states with $\lambda > 1/\sqrt2\approx 0.7071$ violate the Clauser, Horne, Shimony and Holt (CHSH) inequality\cite{CHSH}  - the simplest Bell inequality. It is not known, though, whether the states with $0.6595 \leq \lambda \leq 1/\sqrt2$ violate any Bell inequalities. Recently, the existence of a Bell inequality (which requires at least 465 settings for each party, in contrast to 2 in the case of CHSH inequality) was demonstrated, which is violated for $\lambda \geq 0.7056$.\cite{Ver}

This discussion of Werner states demonstrates that entanglement detection is not equivalent to a violation of Bell's inequality. The former relies on the quantum description of nature, while the latter for example excludes LHV models and probes the very core of quantum mechanics.

\section{Quantum description of ferromagnetic detectors} \label{det}

We assume that the transport channels and the ferromagnetic detectors carry quasi-particles described as single electron spins.
First we aim to detect the projection of an arbitrary spin state on $\ket{\uparrow}_{\vec{n}}$ and $\ket{\downarrow}_{\vec{n}}$, determined by the quantization axis $\vec{n}$ defined by the magnetization of the detectors. In ideal ferromagnets the electric current is fully spin-polarized, which transforms the spin information into the charge information (electric current) by the spin-dependent resistance of the device. The latter can be measured with high precision. Using such detectors we can perform von Neumann measurements described by projection operators
\be
\begin{aligned}
P_\uparrow &= \proj{\uparrow}_{\vec{n}} &\text{ and}\\
P_\downarrow &= \proj{\downarrow}_{\vec{n}}.&
\end{aligned}
\ee
The probability of obtaining a result $i\in \{\uparrow, \downarrow \}_{\vec{n}}$ in a measurement on the state $\rho$ is given by $p_i= \textrm{Tr}(P_i \rho)$ in this ideal case.

However, with non-ideal ferromagnetic detectors with a finite density of states for both spin components at the Fermi surface, an electron in an eigenstate, e.g. $\ket{\uparrow}_{\vec{n}}$, can enter the contact with opposite majority spins and be detected wrongly as $\ket{\downarrow}_{\vec{n}}$ with nonzero probability. Such a measurement process is commonly described by a positive operator valued measure (POVM).\cite{QCQI} In general, a POVM is a set of operators $M_i$ that are positive and fulfill the completeness relation $\sum M_i = I$, with $I$ the identity operator. The elements of the POVM describing inefficient detectors can be chosen as:
\be
\begin{aligned}
M_\uparrow = \Gamma_{+}\proj{\uparrow}_{\vec{n}} + \Gamma_{-}\proj{\downarrow}_{\vec{n}}, \\
M_\downarrow = \Gamma_{+}\proj{\downarrow}_{\vec{n}} + \Gamma_{-}\proj{\uparrow}_{\vec{n}},
\end{aligned}
\ee
with $\Gamma_{+} + \Gamma_{-} = 1$.
It is intuitive to relate $\Gamma_{+}$ and $\Gamma_{-}$ to the spin-dependent density of states $ \nu_{\sigma} $ of a ferromagnetic detector by $\Gamma_{+} = \nu_{\uparrow}/( \nu_{\uparrow} + \nu_{\downarrow} ) $ and $\Gamma_{-} = \nu_{\downarrow}/( \nu_{\uparrow} + \nu_{\downarrow} ) $.
For later use it is more convenient to express the $\Gamma_{\pm}$ in terms of the spin polarization of the ferromagnetic material $\eta$: $ \Gamma_{\pm} = (1 \pm \eta)/2 $, which yields
\ben
\ba
M_\uparrow = \frac{1}{2}(1+\eta)\proj{\uparrow}_{\vec{n}} + \frac{1}{2}(1-\eta)\proj{\downarrow}_{\vec{n}}, \\
M_\downarrow = \frac{1}{2}(1+\eta)\proj{\downarrow}_{\vec{n}} + \frac{1}{2}(1-\eta)\proj{\uparrow}_{\vec{n}},
\ea
\label{emy}
\een
where the spin polarization $\eta \in [0,1]$ is also a measure for the efficiency of the detectors. Each spin state is detected with the probability $\tilde{p}_i= \textrm{Tr}(M_i \rho)$. We note that for $\eta = 1$ one recovers the projective measurements.

The measurement of an electron spin along a given direction $\vec{n}$ with an ideal detector ($\eta = 1$) can be written as the expectation value of the operator
\ben
\hat{\sigma}_{\vec{n}} = P_{\uparrow} - P_{\downarrow} \; ,
\een
with $\hat{\sigma}_{\vec{n}} = \vec{\sigma} \cdot\vec{n}$ and $\vec{\sigma}$ the vector containing the Pauli matrices. A measurement with non-ideal detectors ($\eta < 1$) is equivalent to measuring the observable $\hat{\sigma}^{(\eta)}_{\vec{n}}$ with ideal detectors, where
\ben
\hat{\sigma}^{(\eta)}_{\vec{n}} = M_{\uparrow} - M_{\downarrow}.
\een
From (\ref{emy}) one directly finds $\hat{\sigma}^{(\eta)}_{\vec{n}} = \eta \hat{\sigma}_{\vec{n}}$, which underlines the importance of the spin polarization of the ferromagnetic contact $\eta$ as the efficiency of the detection process.


\section{Equivalence of non-ideal measurements and a depolarizing channel} \label{dep}

In the previous section we have constructed a spin measurement operator for non-ideal ferromagnetic detectors and an input state $\rho$. Now we demonstrate that the use of this measurement operator is equivalent to a specific depolarizing operator $\mathcal{E}(\rho)$ acting on state $\rho$ in the transport channels, and the subsequent detection by ideal ($\eta=1$) ferromagnetic contacts. This view is depicted schematically in Fig. \ref{setup}. First, we define the depolarizing channel $\mathcal{E}$ by
\ben
\mathcal{E}(\rho)=\eta\rho + \frac{1}{2}(1-\eta)I,
\label{depolarization}
\een
and show that for any observable $\mathcal{O}$ the following expression holds:
\ben\label{dual}
\textrm{Tr}\left( \mathcal{O} \mathcal{E}(\rho) \right) = \textrm{Tr}\left( \mathcal{E}(\mathcal{O}) \rho \right).
\label{epsilon_equivalence}
\een
To proof this equality, we reparametrize $\mathcal{E}(\rho)$ with $\eta'= \frac{1+3\eta}{4}$ and the Pauli matrices $\op{i}$,
\ben
\mathcal{E}(\rho)=\eta'\rho + \frac{1-\eta'}{3}(\hat{\sigma}_x \rho \hat{\sigma}_x + \hat{\sigma}_y \rho \hat{\sigma}_y + \hat{\sigma}_z \rho \hat{\sigma}_z).
\een
Evaluating the trace yields
\ben
\lefteqn{\textrm{Tr}\left( \mathcal{O} \mathcal{E}(\rho) \right) = } \nonumber \\
&=&\textrm{Tr}\left(\mathcal{O}\eta'\rho + \frac{1-\eta'}{3}(\mathcal{O}\hat{\sigma}_x \rho \hat{\sigma}_x + \mathcal{O}\hat{\sigma}_y \rho \hat{\sigma}_y + \mathcal{O}\hat{\sigma}_z \rho \hat{\sigma}_z) \right)  \nonumber \\
&=&\textrm{Tr}\left( \eta' \mathcal{O}\rho + \frac{1-\eta'}{3}(\hat{\sigma}_x \mathcal{O}\hat{\sigma}_x \rho  + \hat{\sigma}_y \mathcal{O}\hat{\sigma}_y \rho  + \hat{\sigma}_z \mathcal{O}\hat{\sigma}_z \rho ) \right)  \nonumber \\
&=& \textrm{Tr}\left( \mathcal{E}(\mathcal{O}) \rho \right),
\een
where in the third line we used the linearity of the trace and its invariance under cyclic permutations of the inner matrices.
In particular, one observes that:
\ben
\mathcal{E}(P_i)=M_i,
\een
which, together with Eq.~\eqref{epsilon_equivalence}, shows the equivalence of non-ideal detectors and a depolarization mechanism of the form of Eq.~\eqref{depolarization} with ideal detectors:
\ben
\textrm{Tr}\left( P_i \mathcal{E}(\rho) \right) = \textrm{Tr}\left( M_i \rho \right).
\een


\section{Entanglement witnesses} \label{s:EW}

\subsection{Optimal entanglement witness}

A quantum state $\rho_{AB}$ is entangled, if it is not separable, i.e. if it cannot be written as a convex combination of product states in any basis:
\ben
\rho^{sep}_{AB} = \sum_i q_i \proj{\psi_i}_A \otimes \proj{\phi_i}_B,
\een
with $q_i \geq 0$ and $\sum_i q_i=1$.

The necessary and sufficient condition for separability is limited to density matrices on $\mathcal{H}=\mathcal{C}^2 \otimes \mathcal{C}^2$ and $\mathcal{H}=\mathcal{C}^2 \otimes \mathcal{C}^3$.\cite{Hor.sep} The former is the case for the problems discussed here.
In particular, a state $\rho_{AB}$ is separable if and only if all eigenvalues of the partial transpose $\rho^{T_B}_{AB}$ applied on one subsystem $B$ are positive or zero. The partial transpose is defined by
\ben
\rho^{T_B}_{AB} &=& \left( \sum_{i,j,k,l} c_{ijkl} \ket{i}\bra{j}_A\otimes \ket{k}\bra{l}_B \right)^{T_B} \nonumber \\
            &:=& \sum_{i,j,k,l} c_{ijkl} \ket{i}\bra{j}_A\otimes \ket{l}\bra{k}_B  \\
             &\equiv&  \sum_{i,j,k,l} c_{ijlk} \ket{i}\bra{j}_A\otimes \ket{k}\bra{l}_B. \nonumber
\een

An elegant way of detecting entanglement (not only restricted to low dimensional systems) are entanglement witnesses (EWs).\cite{Hor.sep,Guhne} An EW is a hermitian operator (observable) $\mathcal{W}$ with expectation values $\langle \mathcal{W} \rangle <0$ for at least one entangled state and $\langle \mathcal{W} \rangle \geq0$ for all separable states.

For a given class of states it is possible to construct an EW as\cite{optW}
\ben
\mathcal{W}=\proj{\varphi}^{T_B}_{AB},
\een
where $\ket{\varphi}_{AB}$ is the normalized eigenvector corresponding to a negative eigenvalue of $\rho^{T_B}_{AB}$.
Assuming that we aim to detect maximally entangled singlet states, we insert $\rho_{AB}=\proj{\Psi^-}_{AB}$, which yields the corresponding EW operator
\ben
\mathcal{W} = I - 2\proj{\Psi^-}_{AB}.
\een
We note that this construction can be applied to any other entangled state expected to be generated in an experiment.
$\mathcal{W}$ can be decomposed into spin operators that can be measured locally:
\ben\label{W}
\mathcal{W} =\frac12 ( I + \hat{\sigma}_x \otimes \hat{\sigma}_x + \hat{\sigma}_y \otimes \hat{\sigma}_y + \hat{\sigma}_z \otimes \hat{\sigma}_z).
\een

Now we can calculate the minimum detector efficiency required to witness the entanglement of a maximally entangled input state, here $\proj{\Psi^-}_{AB}$, using the EW operator in Eq.~\eqref{W}.
Since the two particles are measured separately, each by non-ideal detectors with the same efficiency $\eta$, we can view the inefficient detection as each particle of the input state passing an independent depolarizing channel, which results in a Werner state:
\ben\label{Werner}
\mathcal{E}_A\otimes\mathcal{E}_B(\proj{\Psi^-}_{AB})&=& \eta^2 \proj{\Psi^-}_{AB} + (1-\eta^2)\frac{I}{4} \nonumber \\
&\equiv& \rho^{W}_{AB}(\eta^2).
\een
This state is then measured by ideal detectors with an expectation value
\ben
\langle \mathcal{W} \rangle \equiv \textrm{Tr} (\mathcal{W} \rho^{W}_{AB}(\eta^2)) = \frac{1}{2}(1-3\eta^2),
\een
which is negative for spin polarizations
\ben\label{EW1}
\eta > \frac{1}{\sqrt3} \approx 0.58.
\een
This value is a lower bound to the spin polarization of ferromagnetic detectors to detect the entanglement of maximally entangled electron pairs.

These calculations were made using the picture of a noisy transport channel, which is convenient to obtain the minimum efficiency. One obtains the same result using the observable
\ben \label{Weta}
\lefteqn{\mathcal{W}^{(\eta)} = } \nonumber \\
&&\frac12 ( I + \hat{\sigma}^{(\eta)}_x
\otimes \hat{\sigma}^{(\eta)}_x + \hat{\sigma}^{(\eta)}_y \otimes \hat{\sigma}^{(\eta)}_y + \hat{\sigma}^{(\eta)}_z \otimes \hat{\sigma}^{(\eta)}_z),
\een
on the singlet state of Eq.~\eqref{singlet}, in accordance to Eq.~\eqref{dual}.
We note that the spin correlators in Eq.~\eqref{Weta} can be expressed using the experimentally accessible electric current correlators, which we demonstrate in detail in Appendix~\ref{app1}.

\subsection{Entanglement witness with a reduced number of correlators}

To measure the expectation value of the EW in Eq.~\eqref{W} is a formidable experimental challenge, since it requires three orthogonal magnetization directions accessible in the same device.
For this reason we now make use of a reduced EW that contains only two correlators,\cite{otfried,otfr2,otfr3} e.g.
\ben\label{V}
\mathcal{V}_{xy} = \frac12 ( I + \hat{\sigma}_x \otimes \hat{\sigma}_x +  \hat{\sigma}_y \otimes \hat{\sigma}_y).
\een
(In Appendix~\ref{app} we give a simple argument showing that $\langle \mathcal{V}_{xy} \rangle \geq 0$ for all separable states, whereas for a maximally entangled state we have $\langle \mathcal{V}_{xy} \rangle < 0$.) 
In contrast to $\mathcal{W}$, $\mathcal{V}_{xy}$ allows us to detect entanglement by measuring spins only along two orthogonal axes, for example in the substrate plane of a device.

Again, we calculate the minimum requirement for the spin polarization in the leads to detect entanglement using $\mathcal{V}_{xy}$. The expectation value for the noisy input state in Eq.~\eqref{Werner} is
\ben
\langle \mathcal{V}_{xy} \rangle = \textrm{Tr} (\mathcal{V}_{xy} \rho^{W}_{AB}(\eta^2)) =  \frac{1}{2}(1-2\eta^2),
\een
which is negative for the spin polarization
\ben\label{EW2}
\eta > \frac{1}{\sqrt2} \approx 0.71.
\een
For isotropic systems this condition also holds for measurement along the other axes with the EWs
\ben
\mathcal{V}_{xz} = \frac12 ( I + \hat{\sigma}_x \otimes \hat{\sigma}_x +  \hat{\sigma}_z \otimes \hat{\sigma}_z) \; ,\\
\mathcal{V}_{yz} = \frac12 ( I + \hat{\sigma}_y \otimes \hat{\sigma}_y +  \hat{\sigma}_z \otimes \hat{\sigma}_z) \; .
\een
As we can see, the reduction of the number of measured correlators necessitates the use of more efficient detectors, i.e. materials with larger spin polarization.


\section{Violation of Bell's inequalities} \label{bell}

In this section we derive the minimum detection efficiency for test the Bell's inequality in the form of CHSH inequality when using ferromagnetic detectors for electron spins\cite{Kawabata_JPSJ70_2001} to compare with the minimum efficiencies for the EWs derived above.

First, we review the CHSH version of Bell's inequality.\cite{CHSH} Let us assume that there is a local hidden variable description of the system with a set of variables that completely determine the outcomes of all measurements performed by two spatially separated parties. Each party measures one of the two possible dichotomic observables $\alpha_1, \alpha_2, \beta_1$ and $\beta_2$ that can take on the values $\pm1$, e.g. the spin projection along a given axis. Since any measurement outcome is determined by the local hidden variables, either $\alpha_1+\alpha_2=0$ and $\alpha_1-\alpha_2=\pm2$, or $\alpha_1+\alpha_2=\pm2$ and $\alpha_1-\alpha_2=0$ hold, so that for each run of the experiment one obtains
\ben
(\alpha_1+\alpha_2)\beta_1+(\alpha_1-\alpha_2)\beta_2 = \pm2.
\een
Hence, on average
\ben
|\langle\alpha_1\beta_1\rangle+\langle\alpha_2\beta_1\rangle+\langle\alpha_1\beta_2\rangle-\langle\alpha_2\beta_2\rangle| \leq 2,
\een
which is CHSH inequality.
However, in a quantum mechanical description the maximum of this expression is $2\sqrt2$ so that the bound on local hidden variable theories can be violated.

Consider now an operator associated with the CHSH inequality of the form
\ben
\mathcal{B}_{CHSH} = \vec{a}\cdot\hat{\vec{\sigma}} \otimes (\vec{b} + \vec{d})\cdot\hat{\vec{\sigma}} +
                     \vec{c}\cdot\hat{\vec{\sigma}} \otimes (\vec{b} - \vec{d})\cdot\hat{\vec{\sigma}},
\een
where $\vec{a}$, $\vec{b}$, $\vec{c}$ and $\vec{d}$ are arbitrary unit vectors. The corresponding CHSH inequality is violated if for a given state we have $\langle \mathcal{B}_{CHSH} \rangle >2$. The maximum violation $\langle \mathcal{B}_{CHSH} \rangle =2\sqrt2$ occurs for a singlet state and the measurement directions
\ben
\ba
\vec{a} &=& (1,0,0),  \quad
\vec{b} &=& \frac{1}{\sqrt2}(1,0,1),  \\
\vec{c} &=& (0,0,1),  \quad
\vec{d} &=& \frac{1}{\sqrt2}(1,0,-1).
\ea
\label{set}
\een

More generally, a two-qubit state $\rho_{AB}$ violates the CHSH inequality if\cite{Hor.CHSH}
\be\label{warunek}
\sqrt{\lambda_1 + \lambda_2} > 1,
\ee
where $\lambda_1 \geq \lambda_2 \geq \lambda_3$ are the ordered eigenvalues of $R^T R$ with $R_{ij} = \textrm{Tr}[(\op{i} \otimes \op{j}) \rho_{AB}]$.

By choosing $\mathcal{O}=\mathcal{B}_{CHSH}$ in Eq.~\eqref{dual} the measurement of the maximally entangled state with  non-ideal detectors can be expressed by the measurement of the Werner state in Eq.~\eqref{Werner} with an ideal measurement apparatus. A straightforward calculation shows that for this state all eigenvalues of $R^T R$ are $\lambda_i = \eta^4$. By inserting these values into Eq.~\eqref{warunek} one finds that a violation of the CHSH inequality is obtained for spin polarizations or detector efficiencies
\ben
\eta > \frac{1}{\sqrt[4]{2}} \approx 0.84 \;,
\een
in accordance with Ref. [\onlinecite{Kawabata_JPSJ70_2001}].
This minimal polarization is significantly larger than that for the EWs suggested in Eqs.~\eqref{EW1} and \eqref{EW2}. In other words, the requirements for testing Bell's inequalities with ferromagnetic detectors are much stronger than for detecting entanglement with the EWs.


\section{Quantum cryptography with non-ideal detectors} \label{crypto}

One of the most profoundly explored quantum information tasks is quantum cryptography (QC),\cite{crypto} i.e. a protocol that converts correlations obtained from pairs of entangled qubits into secret bits shared by two parties. The single qubit measurements necessary in such protocols can introduce errors, especially when performed with ferromagnetic detectors.

As an example we determine the minimum required detector efficiency for \textit{SIngle-copy Measurement and ClAssical Processing} (SIMCAP) protocols. Such protocols are of special interest when using ferromagnetic detectors as they do not comprise coherent quantum operations, and thus may be experimentally accessible with present day technology.

The basics of SIMCAP key distillation protocols can be described as follows.\cite{acin} Suppose that two parties share $M$ copies of a known state $\rho_{AB}$. In the first step of the protocol, each party performs a filtering process,\cite{filt} so that if it succeeds the resulting state is diagonal in the Bell basis\footnote{The Bell basis is defined by: $\ket{\Psi^\pm} = \frac{1}{\sqrt2}(\ket{\uparrow\downarrow}\pm\ket{\downarrow\uparrow})$, $\ket{\Phi^\pm} = \frac{1}{\sqrt2}(\ket{\uparrow\uparrow}\pm\ket{\downarrow\downarrow})$.}
\ben
\lefteqn{\rho_{AB} \rightarrow  } \nonumber \\
&  \lambda_1 \proj{\Psi^-}_{AB} + \lambda_2 \proj{\Psi^+}_{AB} + \\
& + \lambda_3 \proj{\Phi^-}_{AB} + \lambda_4 \proj{\Phi^+}_{AB} \; , \nonumber
\een
with $\lambda_i \geq 0$ and $\sum_i \lambda_i = 1$.
In the next step each party performs a local unitary ordering transformation with the result that
\ben
\lambda_1 &=& \max \lambda_i, \\
\lambda_2 &=& \min \lambda_i.
\een
After this, a measurement in the basis $\{ \ket{\uparrow}, \ket{\downarrow} \}$ is performed on each qubit and, depending on the outcomes, each party assigns a 0 or 1 to each measurement.
The two parties then share a list of partially correlated bits, which in turn should be processed in a procedure of \textit{advantage distillation}\cite{AD} and \textit{one-way key extraction},\cite{DW} which are performed with the use of possibly insecure but authenticated public classical communication channels.
All these processes are classical procedures that do not require any quantum operations.

With such a protocol it is possible to generate a cryptographic key that is secure against any eavesdropping, provided that\cite{acin}
\ben
\lambda^2_1 + \lambda^2_2 > \frac12 (\lambda_1 + \lambda_2).
\label{SIMCAP_condition}
\een
To obtain the minimum detector efficiency in such a protocol, we assume that we perform the cryptographic protocol given some large number of copies of the state $\proj{\Psi^-}$ and using non-ideal ferromagnetic detectors ($\eta <1$). The problem can again be viewed as using ideal detectors, but performing measurements on a state given in Eq.~\eqref{Werner}. For this we find
\be
\ba
\lambda_1 = \frac14 (3 \eta^2 + 1) \; , \\
\lambda_2 = \frac14 (1 - \eta^2) \; ,
\ea
\ee
which, when inserted into the condition Eq.~\eqref{SIMCAP_condition}, yields the minimal detector efficiency for a SIMCAP protocol
\ben
\eta > \frac{1}{\sqrt[4]{5}} \approx 0.67 \; .
\een
{\ul
This requirement is stronger than for the detection of entanglement, but weaker than for testing Bell's inequality.


\section{Non-ideal entanglement sources} \label{noisy}

In the above analysis of entanglement detection using EWs or Bell operators, we studied pure Bell states in Eq.~\eqref{singlet} as input states. Although high-efficiency sources of entangled spins might be possible,\cite{Schindele_Baumgartner_PRL109_2012} one should investigate the impact of a non-ideal entanglement source. To simulate this, we assume that a maximally mixed component (white noise) is added to the maximally entangled Bell state, which yields a Werner state as in Eq.~\eqref{mixed} with the visibility parameter $\lambda$.
Taking into account the non-ideal detectors in the picture of a quantum depolarizing channel with ideal detectors, one finds directly
\ben\label{noisyWerner}
\mathcal{E}_A\otimes\mathcal{E}_B(\rho^{W}_{AB}(\lambda)) &=& \lambda\eta^2 \proj{\Psi^-}_{AB} + (1-\lambda\eta^2)\frac{I}{4} \nonumber \\
&\equiv& \rho^{W}_{AB}(\lambda\eta^2),
\een
i.e. a Werner state with the visibility parameter $\lambda\eta^2$.
The analysis is then the same as for the state in Eq.~\eqref{Werner}, so that we find the expectation values for the EWs as
\ben
\langle \mathcal{W} \rangle \equiv \textrm{Tr} (\mathcal{W} \rho^{W}_{AB}(\lambda\eta^2)) = \frac{1}{2}(1-3\lambda\eta^2), \\
\langle \mathcal{V}_{xy} \rangle \equiv \textrm{Tr} (\mathcal{V}_{xy} \rho^{W}_{AB}(\lambda\eta^2)) = \frac{1}{2}(1-2\lambda\eta^2),
\een
and $\langle \mathcal{V}_{xy} \rangle = \langle \mathcal{V}_{xz} \rangle = \langle \mathcal{V}_{yz} \rangle $.
Similarly, the condition for the violation of CHSH inequality reads
\ben
2\lambda^2\eta^4 >1.
\een

In Fig.~\ref{Wno} the boundaries for entanglement detection are plotted for the different EWs as a function of the detector efficiency or spin polarization $\eta$ and the visibility parameter $\lambda$ of the Werner state of the emitted electron pair of the entanglement source. It is easy to understand that for larger noise levels (smaller visibility and reduced entanglement) better detectors are required to perform the entanglement detection.

\begin{figure}
  \includegraphics[width=0.4\textwidth]{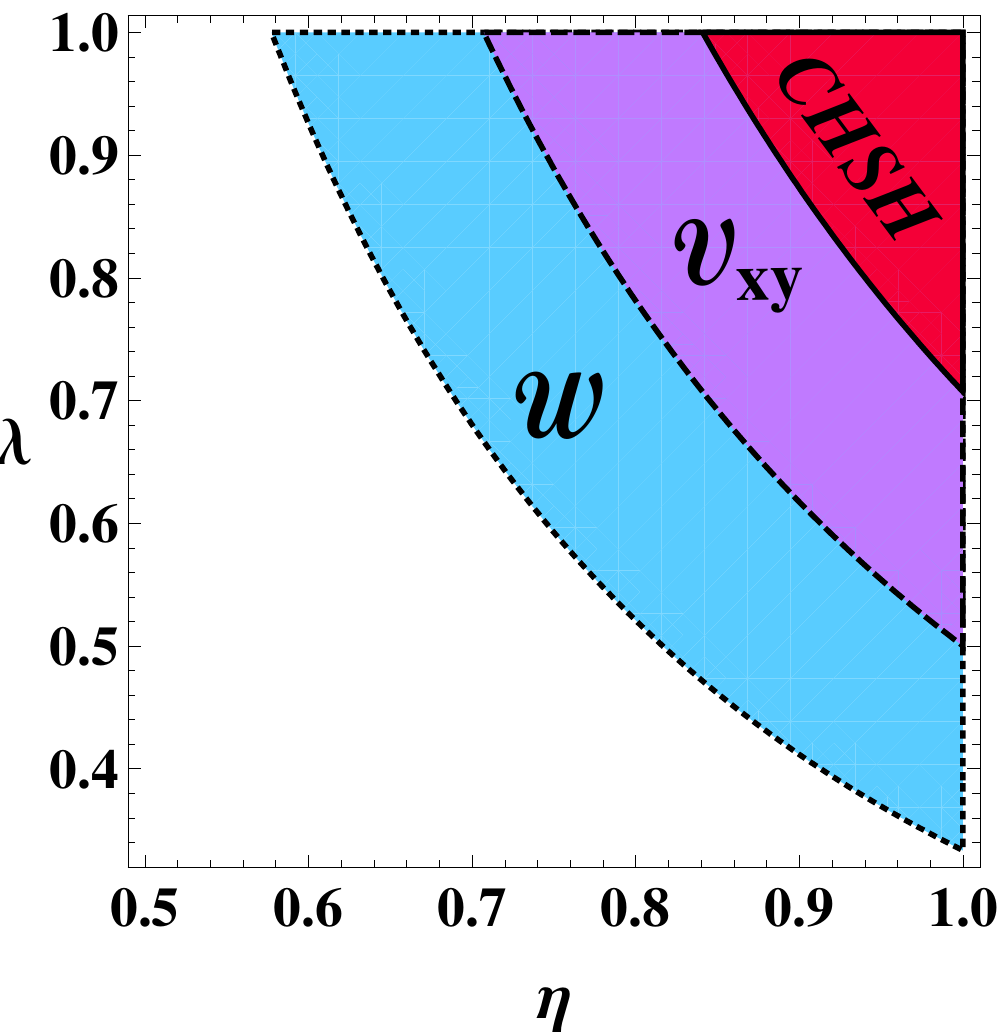}
  \caption{Map of possible entanglement detection for different source visibility $\lambda$ and spin polarization $\eta$ for the EWs described in the text. $\langle \mathcal{W} \rangle <0$ (above dotted line) with three correlators, and $\langle \mathcal{V}_{xy} \rangle <0$ (above dashed line) with two correlators allow entanglement detection, as well as the region for which the CHSH inequality is violated (above solid line).}
  \label{Wno}
\end{figure}


\section{Spin relaxation and dephasing} \label{deph}

The effects of relaxation and dephasing of the electron spins in real transport channels are not described by Werner states, but cannot be neglected in our investigation. In the following we adopt a phenomenological model for spin decoherence\cite{rapid} used recently to obtain limits for the relaxation and the dephasing times in entanglement detection with CHSH inequality.\cite{ita} Here we use this model to find the conditions for detecting entanglement using the presented EWs and for finding the optimal angles in $\mathcal{B}_{CHSH}$ also for noisy states of the form in Eq.~\eqref{mixed}, all using non-ideal detectors.

Spin decoherence can be described by the channel $\Lambda$ that acts on each particle's state as:
\ben
\ba
\proj{\uparrow} \overset{\Lambda}{\longrightarrow} & \,\,  \frac{1+\omega_1}{2}\proj{\uparrow} + \frac{1-\omega_1}{2}\proj{\downarrow}, \\
\proj{\downarrow} \overset{\Lambda}{\longrightarrow} & \,\, \frac{1+\omega_1}{2}\proj{\downarrow} + \frac{1-\omega_1}{2}\proj{\uparrow}, \\
\ket{\uparrow}\bra{\downarrow} \overset{\Lambda}{\longrightarrow} & \,\, \omega_2 \ket{\uparrow}\bra{\downarrow}, \\
\ket{\downarrow}\bra{\uparrow} \overset{\Lambda}{\longrightarrow} & \,\, \omega_2 \ket{\downarrow}\bra{\uparrow},
\ea
\label{dec}
\een
where $\omega_1 = \exp(-t_L / T_1)$, $\omega_2 = \exp(-t_L / T_2)$, and $T_1$ and $T_2$ are the spin relaxation and the spin dephasing times, respectively. $t_L = L / v_F$ is the ballistic transmission time though the transport channel, with $L$ the length of the channel and $v_F$ the Fermi velocity. We note that in this model the dephasing time includes pure dephasing as well as dephasing due to relaxation process, which in general are related by
\ben
\frac{1}{T_2} = \frac{1}{2 T_1} +\frac{1}{T^*_2},
\een
where $T^*_2$ is the time-scale for pure dephasing.
In particular, for pure relaxation we have $T_2=2T_1$.


Assuming that decoherence and relaxation are independent in each channel, the pure singlet state evolves into
\ben\label{decoheredstate}
\rho^{decoh}_{AB} = \Lambda_A \otimes \Lambda_B (\proj{\Psi^-}_{AB}) \; ,
\een
which is not a Werner state, but still diagonal in the Bell basis.

The expectation values of the EWs can then be found as
\ben
\langle \mathcal{W} \rangle \equiv \textrm{Tr} (\mathcal{W} \rho^{decoh}_{AB}) &=& \frac12 (1- \omega^2_1 - 2\omega^2_2), \\
\langle \mathcal{V}_{xy} \rangle \equiv \textrm{Tr} (\mathcal{V}_{xy} \rho^{decoh}_{AB}) &=& \frac12 - \omega^2_2, \\
\langle \mathcal{V}_{xz} \rangle \equiv \textrm{Tr} (\mathcal{V}_{xz} \rho^{decoh}_{AB}) &=& \frac12 (1- \omega^2_1 - \omega^2_2), \\
\langle \mathcal{V}_{yz} \rangle \equiv \textrm{Tr} (\mathcal{V}_{yz} \rho^{decoh}_{AB}) &=& \frac12 (1- \omega^2_1 - \omega^2_2).
\een
The ranges of $t_L / T_1$ and $t_L / T_2$ for which these mean values are negative are presented in Fig.~\ref{rel}.

\begin{figure}
  \includegraphics[width=0.4\textwidth]{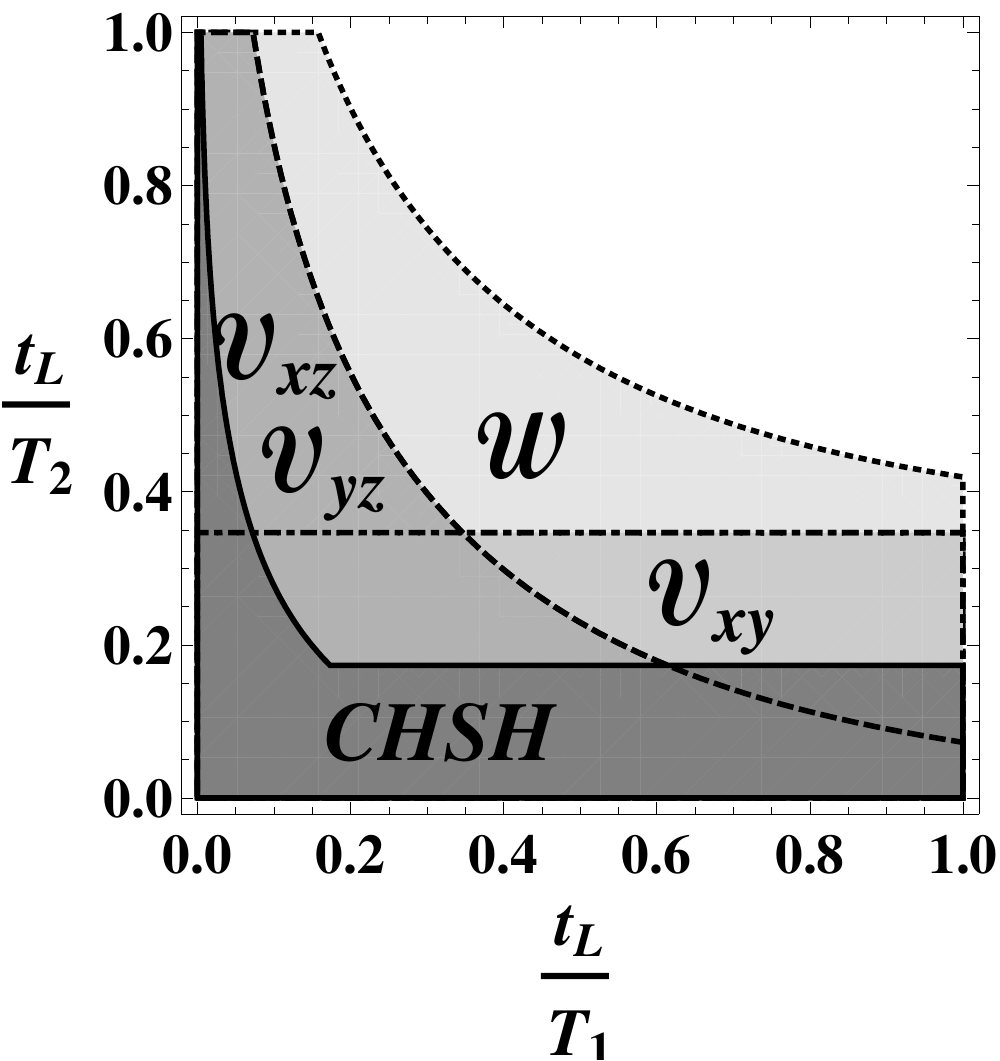}
  \caption{The region of the transmission time $ t_L $ relative to the spin relaxation time, $t_L / T_1$ and and spin dephasing time, $t_L / T_2$, for which the entanglement witnesses $\langle \mathcal{W} \rangle <0$ (below dotted line), $\langle \mathcal{V}_{xz}\rangle, \langle\mathcal{V}_{yz} \rangle <0$ (below dashed line), $\langle \mathcal{V}_{xy} \rangle <0$ (below dash-dotted line) and the region for the parameters required for violation of the CHSH inequality (below solid line). Only the singlet state has been considered.}
  \label{rel}
\end{figure}

It should be stressed that, although the mean values of different EWs calculated for the singlet state $\proj{\Psi^-}$ were equal,
the decoherence described by Eq.~\eqref{dec} breaks the rotational symmetry of $\proj{\Psi^-}$, which results in different expectation values of $\mathcal{V}_{ij}$ measured with correlators in different planes.

The condition for the violation of the CHSH inequality reads
\ben
\max \{ 2\omega^4_2, \omega^4_1 + \omega^4_2 \} > 1,
\een
which suggests a much broader parameter range to detect entanglement by means of violation of the CHSH inequality than found in earlier studies.\cite{ita} Here we use a different method\cite{Hor.CHSH} to obtain the optimal set of angles that results in the maximum values of $\langle \mathcal{B}_{CHSH} \rangle$ for the decohering state in Eq.~\eqref{decoheredstate}. For $T_1<T_2$ one finds
\ben
\ba
\vec{a} &= (1,0,0),  \\
\vec{b} &= \frac{1}{\sqrt2}(-1,-1,0),  \\
\vec{c} &= (0,1,0),  \\
\vec{d} &= \frac{1}{\sqrt2}(-1,1,0) ,
\ea
\label{set2}
\een
while for $T_1>T_2$:
\ben
\ba
\vec{a} &= (0,0,1),  \\
\vec{b} &= (-\sin\theta,0,-\cos\theta),  \\
\vec{c} &= (1,0,0),  \\
\vec{d} &= (\sin\theta,0,-\cos\theta) ,
\ea
\label{set3}
\een
where $\sin\theta = 1/\sqrt{1+\exp(-4\Delta)}$, $\cos\theta = 1/\sqrt{1+\exp(4\Delta)}$ and $\Delta = t_L(\frac{1}{T_1}-\frac{1}{T_2})$.
%
In Fig.~\ref{rel} we plot the boundaries for the violation of the CHSH inequality as a function of the parameters $t_L / T_1$ and $t_L / T_2$. The kink in this curve is due to the change of the optimal detector angles between Eq.~\eqref{set2} and Eq.~\eqref{set3}.

The analysis can be further generalized to Werner states given by Eq.~\eqref{Werner} as a result of white noise and non-ideal detectors. Again, in the spirit of Fig. \ref{setup}, this case can be viewed from two perspectives: an incoming initial state $\rho^W_{AB}(\lambda)$ propagates through the transport channels and is affected by relaxation and decoherence ($\Lambda_A \otimes \Lambda_B$), and the resulting state is measured with non-ideal ferromagnetic detectors. In turn, one can also perceive the problem as a spin measurement with ideal detectors on a state
\ben
\lefteqn{\mathcal{E}_A \otimes \mathcal{E}_B \left( \Lambda_A \otimes \Lambda_B (\rho^W_{AB}(\lambda))   \right) =} \nonumber \\
&=& \Lambda_A \otimes \Lambda_B \left( \mathcal{E}_A \otimes \mathcal{E}_B (\rho^W_{AB}(\lambda))   \right) \nonumber \\
&=& \Lambda_A \otimes \Lambda_B (\rho^W_{AB}(\lambda \eta^2)) \nonumber \\
&\equiv&  \rho^{decoh}_{AB}(\lambda \eta^2),
\een
where the first equality can be found by direct calculation. From this expression we can calculate the expectation values of the different EWs:
\ben
\label{set4.1}
\langle \mathcal{W} \rangle       &=& \frac12 [ 1- \lambda\eta^2(\omega^2_1 + 2\omega^2_2)], \\
\label{set4.2}
\langle \mathcal{V}_{xy} \rangle  &=& \frac12 - \lambda\eta^2\omega^2_2, \\
\label{set4.3}
\langle \mathcal{V}_{xz} \rangle = \langle \mathcal{V}_{yz} \rangle  &=& \frac12 [ 1- \lambda\eta^2(\omega^2_1 + \omega^2_2)].
\een
Entanglement is detected whenever these values are negative. Again we note that the $xy$ plain becomes discriminated due to the specific character of the decoherence process. Namely, we see that due to the absence of the correlator $\hat{\sigma}_z \otimes \hat{\sigma}_z$ in the witness $\mathcal{V}_{xy}$, the effect of relaxation can be measured only with respect to dephasing.
The condition for the violation the CHSH inequality reads
\ben \label{ine}
\max \{ 2\lambda^2\eta^4\omega^4_2, \lambda^2\eta^4 (\omega^4_1 + \omega^4_2) \} > 1,
\een
when using the optimal set of angles given in Eq.~\eqref{set2} and Eq.~\eqref{set3}.

Eqs.~\eqref{set4.1}~-~\eqref{set4.3} and the CHSH inequality in Eq.~\eqref{ine} show that in the presence of spin relaxation and dephasing, ferromagnetic detectors with a larger efficiency, i.e. with a higher spin polarization, are required to demonstrate spin entanglement.


\section{Summary and Discussion} \label{dis}


We have used several entanglement witnesses (EWs)
for detection of the electrons spin states entanglement and discussed the impact of non-ideal ferromagnetic detectors, white noise, spin relaxation and decoherence.
In addition, we compare these results to tests of Bell's inequality and calculate detector efficiency requirement for a quantum cryptography protocol. We can order these different experiments by increasing requirements for the detector efficiencies or spin polarization and propose a "road map" for proving spin entanglement or a violation of Bell inequality with electron spins:

(i) A spin correlation measurement with colinear magnetizations of the detector leads can show
\ben
\langle \hat{\sigma}_z \otimes \hat{\sigma}_z \rangle < 0 \; ,
\een
which is possible for any $ \eta > 0 $. This experiment would prove that two electrons from a source contact have opposite spins.

(ii) Assuming that the system under consideration is rotationally invariant, which is true for all Werner states, we have that $ \langle \hat{\sigma}_z \otimes \hat{\sigma}_z \rangle = \langle \hat{\sigma}_x \otimes \hat{\sigma}_x \rangle = \langle \hat{\sigma}_y \otimes \hat{\sigma}_y \rangle$. Then if a single collinear spin correlator fulfills the inequality
\ben \label{ss}
\langle \hat{\sigma}_z \otimes \hat{\sigma}_z \rangle < -1/3 \; ,
\een
which is technically less demanding than three orthogonal correlators measurements, one could argue that from Eq.~\eqref{ss} we have $ \mathcal{W} <0 $ which requires the spin polarization $ \eta > 1/\sqrt{3} \approx 0.58 $. Note, however, that the original assumption is crucial here: if the state under consideration is e.g. a product state $\ket{\uparrow_z}_A\otimes \ket{\downarrow_z}_B$, then $ \langle \hat{\sigma}_z \otimes \hat{\sigma}_z \rangle = -1$, whereas $ \langle \hat{\sigma}_x \otimes \hat{\sigma}_x \rangle = \langle \hat{\sigma}_y \otimes \hat{\sigma}_y \rangle =0$, and in this case we would not be able to distinguish between the product state and the maximally entangled singlet state.

(iii) Experimentally more demanding is to measure EWs with two collinear in-plane spin correlators such as $ \mathcal{V}_{xy} < 0 $, which requires $ \eta > 1/\sqrt{2} \approx 0.71$.

(iv) More difficult is to test the condition $ \mathcal{W} <0 $ without the assumption of rotational invariance. It requires the measurements of three collinear correlators in the $x$, $y$, and $z$ directions. The condition for the spin polarization $ \eta  > 0.58 $ is less restrictive than for the previous case, but the engineering of the magnetization directions might be less trivial.

(v) In order to perform a quantum cryptographic SIMCAP protocol using pure singlet states, the spin polarization of spin detectors have to be $ \eta > 0.67$. 

(vi) We find that the most difficult task to perform is the violation of Bell's inequalities. It would require in-plane measurements of four noncollinear spin correlators and the largest detector efficiencies, $ \eta > 0.84$.

Note, however, that all the above threshold values apply to measurements on the pure singlet states, whereas for noisy states they are respectively more stringent.


Nanometer scale contacts with spin polarizations larger than 58\% are difficult to fabricate. Ferromagnetic materials used to contact nanostructures,\cite{Tombros_van_der_Molen_van_Wees_PRB73_2006, Aurich_Baumgartner_APL97_2010} like the elements Ni, Co and Fe, or simple alloys, are limited to polarizations in the range of $30-50$\%.\cite{Meservey_Tedrow_PhysRep238_1994, Soulen_Science282_1998, Moodera_AnnuRevMaterSci_1999}
In principle, spin polarization of close to 100\% could be obtained in half-metals, for which several candidates are currently investigated extensively. Up to 90\% polarization was reported for CrO$_2$\cite{Soulen_Science282_1998, Parker_PRL88_2002} which, however, is meta-stable in air and decomposes into other oxides. A series of ternary and quaternary Heusler alloys exhibit polarizations up to 74\%. \cite{Varaprasad_ActaMaterialia_2012_Heusler_alloys_PCAR} The fabrication of a nanoelectronic device with La$_{0.7}$Sr$_{0.3}$MnO$_3$ (LSMO) contacts was already demonstrated, possibly with a polarization on the order of 80\% \cite{Hueso_Fert_Nature445_2007}. Another possible route to spin injection and detection is the use of ferromagnetic insulators as tunnel barriers.\cite{Santos_Moodera_PRB69_2004_EuO} Though the device fabrication with all these compounds is rather involved  and requires a very high degree of material and interface control, the reported polarizations demonstrate the feasibility of detecting entanglement with ferromagnetic detectors.

%

\begin{acknowledgements}

We would like to thank Adam Bednorz, Wolfgang Belzig, Bernd Braunecker, Micha\l{} Horodecki, Ryszard Horodecki, Alfredo Levy Yeyati, Daniel Loss, Gerd Sch\"{o}n, and Antoni W\'{o}jcik for helpful discussions.
AB, CS, DT, and JM were supported by the EU FP7 project SE2ND [271554] and DT and JM by a Polish grant for science for the years 2009-2015.
WK and AG were supported by the Polish Ministry of
Science and Higher Education Grant no. IdP2011 000361.
WK thanks the Foundation of Adam Mickiewicz University in Pozna\'{n} for the support from its
scholarship programme.

\end{acknowledgements}

\appendix

\section{} \label{app1}

To determine the expectation value of the entanglement witnesses one has to measure the spin correlation $ \langle \hat{n}_A \cdot \hat{\sigma} \otimes \hat{n}_B \cdot \hat{\sigma}  \rangle  $. It can be expressed in terms of the correlators $ \langle N_{\hat{n}_A \alpha}(\tau)  N_{\hat{n}_B \beta}(\tau) \rangle $ of electrons number with spin $ \alpha $ ($ \beta $) detected at a time $ \tau $ in the detector $A$ ($B$) that correspond to a coincidence counting in the limit when a single particle per detector arrives at a time $ \tau $.
The spin correlation is given by
\ben
\lefteqn{\langle \hat{n}_A \cdot \hat{\sigma} \otimes \hat{n}_B \cdot \hat{\sigma}  \rangle = } \\
&=& \frac{\langle [N_{\hat{n}_A \uparrow}(\tau)-N_{\hat{n}_A \downarrow}(\tau)][N_{\hat{n}_B \uparrow}(\tau)-N_{\hat{n}_B \downarrow}(\tau)] \rangle}{\langle [N_{\hat{n}_A \uparrow}(\tau) + N_{\hat{n}_A \downarrow}(\tau)][N_{\hat{n}_B \uparrow}(\tau) + N_{\hat{n}_B \downarrow}(\tau)] \rangle} \; . \nonumber
\een
For simplicity we consider the tunneling regime when the current pulses width is smaller than their spacing. As it was demonstrated in Ref.~\cite{current1,current2}, even though it is technically difficult to detect individual current pulses, it is possible to determine the correlators $ \langle N_{\hat{n}_A \alpha}(\tau)  N_{\hat{n}_B \beta}(\tau) \rangle $ by measurement of low frequency current correlations (noise). The correlators can be obtained by integrating the current correlator over time,
\be \label{cor1}
\langle N_{\hat{n}_A \alpha}(\tau)  N_{\hat{n}_B \beta}(\tau) \rangle
=\int_0^{\tau} dt \int_0^{\tau} dt' \langle I_{\hat{n}_A \alpha}(t) I_{\hat{n}_B \beta}(t') \rangle \; ,
\ee
since the number of electrons detected over time $ \tau $ in the detector $A$ is given by
$
\langle N_{\hat{n}_A \alpha}(\tau)  \rangle = \int_0^{\tau} dt \langle I_{\hat{n}_A \alpha}(t) \rangle
$.
The current correlator in Eq.~\eqref{cor1} can be expressed in terms of the zero-frequency current correlator (noise) defined as
\be
S_{\hat{n}_A \alpha , \hat{n}_B \beta}(0) \equiv \int dt \langle \delta I_{\hat{n}_A \alpha}(t) \delta I_{\hat{n}_B \beta}(t') \rangle \; ,
\ee
where
$ \delta I_{\hat{n}_A \alpha}(t) =  I_{\hat{n}_A \alpha}(t) - \langle  I_{\hat{n}_A \alpha}(t) \rangle $.
In the limit when $\tau$ is larger than the current pulses width the correlator form Eq.~\eqref{cor1} is given by
\be
\langle N_{\hat{n}_A \alpha}(\tau)  N_{\hat{n}_B \beta}(\tau) \rangle
= \tau^2 \langle I_{\hat{n}_A \alpha} \rangle \langle I_{\hat{n}_B \beta} \rangle + \tau S_{\hat{n}_A \alpha , \hat{n}_B \beta}(0)
 \; ,
\ee
whereas for a short time $\tau $ in comparison with the current pulses spacing one may neglect the first term. The spin correlator is then given in terms of the low-frequency current correlators (noise),
\begin{widetext}
\be
\langle \hat{n}_A \cdot \hat{\sigma} \otimes \hat{n}_B \cdot \hat{\sigma}  \rangle = \gamma(\tau, e V)
\frac{S_{\hat{n}_A \uparrow , \hat{n}_B \uparrow}(0)
+ S_{\hat{n}_A \downarrow , \hat{n}_B \downarrow}(0)
- S_{\hat{n}_A \uparrow , \hat{n}_B \downarrow}(0)
- S_{\hat{n}_A \downarrow , \hat{n}_B \uparrow}(0)
}
{
S_{\hat{n}_A \uparrow , \hat{n}_B \uparrow}(0)
+ S_{\hat{n}_A \downarrow , \hat{n}_B \downarrow}(0)
+ S_{\hat{n}_A \uparrow , \hat{n}_B \downarrow}(0)
+ S_{\hat{n}_A \downarrow , \hat{n}_B \uparrow}(0)}
\; .
\ee
\end{widetext}

\section{} \label{app}


To show that the expectation value of $\langle \mathcal{W} \rangle=\textrm{Tr}(\mathcal{W} \rho^{sep}_{AB})$ is nonnegative for separable states $\rho^{sep}_{AB}$
\ben
\rho^{sep}_{AB} = \sum_i q_i \proj{\psi_i}_A \otimes \proj{\phi_i}_B,
\een
($\sum_i q_i = 1$, $q_i \geq 0$), by linearity of the trace it suffices to check whether for all product states $\proj{\psi_i}_A \otimes \proj{\phi_i}_B$ we have
\ben\label{VV}
\textrm{Tr}(\mathcal{W} \proj{\psi_i}_A \otimes \proj{\phi_i}_B)\geq 0.
\een

If we set:
\ben\label{sety}
\ba
\ket{\psi}  = \sin\alpha\ket{\uparrow}+e^{i \theta}\cos\alpha\ket{\downarrow},\\
\ket{\phi}  = \sin\beta\ket{\uparrow}+e^{i \phi}\cos\beta\ket{\downarrow},
\ea
\een
we obtain:
\begin{widetext}
\ben
\lefteqn{\textrm{Tr} \left( \frac12 ( I + \hat{\sigma}_x \otimes \hat{\sigma}_x +  \hat{\sigma}_y \otimes \hat{\sigma}_y +   \hat{\sigma}_y \otimes \hat{\sigma}_y) \proj{\psi}_A\otimes\proj{\phi}_B \right) = } \nonumber \\
&=& \frac12 (1+\langle \psi | \hat{\sigma}_x | \psi \rangle \langle \phi | \hat{\sigma}_x | \phi \rangle
+ \langle \psi | \hat{\sigma}_y | \psi \rangle \langle \phi | \hat{\sigma}_y | \phi \rangle + \langle \psi | \hat{\sigma}_z | \psi \rangle \langle \phi | \hat{\sigma}_z | \phi \rangle) \nonumber \\
&=& \sin^2\alpha \sin^2\beta  + \cos^2\alpha \cos^2\beta       +\frac12   \cos(\theta-\phi)\sin2\alpha\sin2\beta \nonumber \\
&\geq& 0,
\een
\end{widetext}
which proves \eqref{VV}.

The same argument applies for the expectation value of $\langle \mathcal{V}_{xy} \rangle=\textrm{Tr}(\mathcal{V}_{xy} \rho^{sep}_{AB})$. Again, for \eqref{sety} we have:
\ben
\lefteqn{\textrm{Tr} \left( \frac12 ( I + \hat{\sigma}_x \otimes \hat{\sigma}_x +  \hat{\sigma}_y \otimes \hat{\sigma}_y) \proj{\psi}_A\otimes\proj{\phi}_B \right) = } \nonumber \\
&=& \frac12 (1+\langle \psi | \hat{\sigma}_x | \psi \rangle \langle \phi | \hat{\sigma}_x | \phi \rangle
+ \langle \psi | \hat{\sigma}_y | \psi \rangle \langle \phi | \hat{\sigma}_y | \phi \rangle) \nonumber \\
&=&  \frac12 (1+\cos(\theta-\phi)\sin2\alpha\sin2\beta) \nonumber \\
&\geq& 0.
\een

The argument for $\mathcal{V}_{xz}$ and $\mathcal{V}_{yz}$ follows similar lines and we obtain:
\ben
\lefteqn{\textrm{Tr}(\mathcal{V}_{xz} \proj{\psi}_A \otimes \proj{\phi}_B) =} \nonumber \\
&=& \frac12 (1+\cos2\alpha\cos2\beta+\cos\theta\cos\phi\sin2\alpha\sin2\beta)  \nonumber \\
&\geq& 0, \\
\lefteqn{\textrm{Tr}(\mathcal{V}_{yz} \proj{\psi}_A \otimes \proj{\phi}_B) =} \nonumber \\
&=& \frac12 (1+\cos2\alpha\cos2\beta+\sin\theta\sin\phi\sin2\alpha\sin2\beta)  \nonumber \\
&\geq& 0,
\een
which proves that $\mathcal{V}_{xy}$, $\mathcal{V}_{xz}$ and $\mathcal{V}_{yz}$ are EWs.

\end{document}